\documentstyle[11pt]{article}

\small
\begin{document}
\begin{titlepage}

\title{
{\bf Isovector response function of hot nuclear matter with Skyrme
interactions } }
\author{  {\bf F\'abio L. Braghin}\thanks{Doctoral fellow of Coordena\c c\~ao
de Aperfei\c coamento de Pessoal de N\'\i vel Superior,Brasil }~,
 {\bf Dominique Vautherin}, \\
{\normalsize
Division de Physique Th\'eorique\thanks{Unit\'e de Recherche des
Universit\'es Paris XI et Paris  VI associ\'ee au CNRS},
 Institut de Physique Nucl\'eaire,}\\
{\normalsize 91406 Orsay Cedex, France }
\\
{\bf Abdellatif  Abada} \\
{\normalsize   Institute for Theoretical Physics, T\"ubingen University, }\\
{\normalsize   Auf der Morgenstelle  14, D-72076  T\"ubingen, Germany}
}

\date{}
\maketitle
\begin{abstract}
We investigate the role of the effective nucleon-nucleon interaction in the
description of giant
dipole resonances in hot nuclei.
For this purpose we calculate the response function of hot
nuclear matter
to a small isovector external perturbation using various effective
Skyrme interactions.
We find that for Skyrme forces with an effective mass close to unity
an undamped zero sound mode occurs at zero temperature.
This mode gives rise in finite nuclei (calculated via the Steinwedel-Jenssen
model) to a resonance whose energy agrees with
the observed value. We find that zero sound disappears at a
temperature of a few MeV, leaving only a broad peak in the dipole strength.
For Skyrme forces with a small value of the effective mass (0.4- 0.5), there is
no zero sound at zero temperature but only a weak peak located too high in
energy.
The strength distribution in this case is nearly independent of temperature and
shows small collective effects.
The relevance of these results for the saturation of photon multiplicities
observed in recent experiments is pointed out.
\end{abstract}

\vskip 0.4cm

\vskip 0.4cm

 UNITU-THEP-7/1995

 IPNO/TH 95-11 \hfill{March 1995}

\end{titlepage}

\renewcommand{\thesection}{\Roman{section}.}
\renewcommand{\theequation}{\arabic{section}.\arabic{equation}}
\newpage
\setcounter{equation}{0}
\section{Introduction}

Giant dipole resonances built on nuclear excited  states have been
the subject of numerous studies since their first observation at the Berkeley
88" cyclotron in 1981 \cite{DRANEW}.
By now a significant amount of information is available concerning
their evolution with increasing excitation energy
(which is mainly stored in
rotational and thermal degrees of freedom)
\cite{GAARDHOJE,TAPS,BRACCO,ALAMAU}. The energy of the
resonance is known to a good accuracy to be nearly independent of the
excitation energy ($E^*$) while
the width increases up to approximatly $E^* \sim $130 MeV in the
case of nuclei with $ A \simeq$110. Beyond this value there are
indications that the width saturates \cite{BRACCO,ALAMAU,LEFAOU}
although some experiments suggest in contrast a steady increase
\cite{YOSHI90,KAG92}.
In the recent experiments performed at
the GANIL facility with a 37 MeV per nucleon Argon beam, the saturation
of the width was observed at excitation energies
greater than about 250 MeV \cite{LEFAOU} together with a saturation of the
photon multiplicity. At such high excitation energies,
it is not clear, however, whether or not thermalization is actually reached.

There are experimental indications that angular momentum contributes
significantly to the width of the resonance
up to the point where scission occurs \cite{BRACCO}.
Although some calculations of this effect \cite{BRO87,ALH90,GAL85} show
a rapid and continuous increase of the width with angular momentum, they
do not seem to account neither for the observed variation at mean and high
excitation energies nor for the saturation.

At zero temperature the dominant contribution to the width
is the spreading over two particle-two hole configurations \cite{ADACHI}.
It shows however a negligible variation with temperature \cite{BORT86,DIDA89}.
In contrast, in Ref. \cite{NICOLE}, a rapid increase of the
width at small temperatures and a saturation beyond $T = 4$ MeV was obtained
by including explicitely
those additional configurations (particle-particle and hole-hole states)
which begin to occur in Random Phase Approximation (RPA) calculations
as temperature rises.

RPA calculations generally give resonance energies in
agreement with each other (and with experimental data). In contrast,
rather different predictions
can be found in the literature about
the width dependence on temperature. Some
RPA results for the giant dipole resonance for  $^{40}$Ca predict a
sharp peak at zero temperature and an increase of the width with temperature
\cite{VAUTHERIN} whereas such is not the case for the self-consistent RPA
calculations of Ref. \cite{SAGAWA}. In this last case, one already has at zero
temperature a fragmented resonance displaying little evolution with
temperature.
In the nuclear matter calculations of Refs.
\cite{BRAVAU,RIKEN} it was found that for specific values of
the particle-hole interaction  strength  a
disappearance of the collective (zero sound type) mode occurs at
a temperature of a few MeV.
This result was however obtained in the case of a schematic Skyrme-type
interaction which contains no momentum dependent terms and therefore
cannot be considered as a reliable effective force.

The purpose of the present paper is to investigate
what can be learned from nuclear matter
calculations of the isovector response function
at finite temperature using more appropriate effective forces. In what follows
we consider the standard effective
Skyrme interactions which have been successful at describing
accurately nuclear ground state properties
while retaining contact with the standard fundamental description of nuclear
matter based on the Brueckner reaction matrix \cite{NEGEVAU}.
Calculations with such Skyrme forces are somewhat tedious.
Nevertheless we will show that analytical formulae can be obtained which are
still quite transparent. These formulae generalize those of Garcia-
Reccio {\it et al.} \cite{GARCIA} for the zero temperature response function.
These formulae will allow us to discuss in rather general terms the evolution
of the collective behaviour of nuclear matter with temperature. In particular
we will see that zero sound is not as robust in nuclear matter as in usual
Fermi liquids and disappears at temperatures of a few MeV for all the standard
Skyrme forces investigated in the present work.

\setcounter{equation}{0}
\section{The response function of hot nuclear matter}

We calculate the response function of  nuclear matter at
finite temperature to an infinitesimal external field of the form
\begin{equation} \label{1}
V_{\rm ext}= \varepsilon \tau_3 e^{-i {\bf q}.{\bf r} }
e^{-i (\omega +i \eta) t},
\end{equation}
where $\tau_3$ is the third isospin Pauli matrix and $\eta$ a vanishingly
small positive number corresponding to an
adiabatic switching of the field from the time $t=-\infty$.
The temporal evolution of the one-body density matrix
$\rho$ is determined by the time-dependent Hartree- Fock equation in
the presence of the perturbation term, i.e.,
\begin{equation} \label{2}
i \hbar \partial_t \rho= [W + V_{\rm ext},  \rho ],
\end{equation}
We consider effective Skyrme forces (see
appendix A for details concerning the Skyrme force).
In this case, the energy density can be written in terms of the one
body density, kinetic energy density $\tau$ and momentum density $ {\bf j}$.

\renewcommand{\thesubsection}{A.}
\subsection{ Construction of the Response Function }

For a small enough external field it is legitimate to linearize the mean
field evolution equation (\ref{2}) around its static solution. This
procedure leads to the following approximate equation in momentum space for
the difference
$\delta \rho = \rho_n - \rho_p$ between the neutron and proton density
matrices:
\begin{equation} \label{5}
\begin{array}{ll}
i \hbar \partial_t \langle {\bf k} | \delta \rho | {\bf k}'\rangle=
&( \epsilon({\bf k})- \epsilon({\bf k}') ) \langle {\bf k}| \delta \rho|
{\bf k}'\rangle  + 2 ( f({\bf k}')-f({\bf k}) ) \langle {\bf k}|
(W_n-W_p)| {\bf k}'\rangle  \\
&+ 4 \varepsilon ( f({\bf k}')-f({\bf k}) )~
\delta({\bf k}'-{\bf k}- {\bf q}) ~e^{-i (\omega +i \eta) t} .
\end{array} \end{equation}
In this equation $ \epsilon ({\bf k}) = \hbar^2 k^2 / 2m^*$  (with $m^*$
being the effective mass; see appendix A) is the kinetic energy of a
single particle state with momentum $ {\bf k}$ in symmetric nuclear matter
and $f({\bf k})$ is the corresponding occupation number
\begin{equation} \label{5a}
\displaystyle
f({\bf k})=1/(1+e^{\beta(\epsilon({\bf k})- \mu )} ),
\end{equation}
where $\beta = 1/T$ is the inverse temperature.
Note that since $\epsilon ({\bf k})$ contains the kinetic energy only, the
quantity $\mu$ in the previous equation is not quite the chemical potential
$ \bar{\mu}$ but rather $ \bar{\mu} - U$,
where $U$ is the mean field defined in appendix A.
{}From here on we will adopt the standard units in which $\hbar = c =1$.

The external field
(\ref{1}) induces  a difference between neutron and proton
density distributions for which we consider the following time-dependent
Ans\"atze suggested by the form of $ V_{\rm ext}$
\begin{equation} \label{12} \begin{array}{ll}
\langle {\bf r}| \delta \rho|{\bf r}\rangle  =
\alpha e^{-i {\bf q}.{\bf r} } e^{-i (\omega +i \eta) t} \\
\langle{\bf r}| \delta \tau|{\bf r}\rangle  =
\beta  e^{-i {\bf q}.{\bf r} } e^{-i (\omega +i \eta) t} \\
\langle{\bf r}| \delta {\bf j}|{\bf r}\rangle =
\gamma {\bf q} e^{-i {\bf q}.{\bf r} } e^{-i (\omega +i \eta) t}
\end{array}
\end{equation}
{}From the definitions of the densities given in appendix A the coefficients
$\alpha$, $\beta$ and $\gamma$ must satisfy
\begin{equation} \begin{array}{ll}
{\displaystyle \left( \alpha , \beta , \gamma \right)  =
\int \frac{{\rm d}^3 k}{ (2\pi)^3} \left( 1 , {\bf k}.({\bf k+q}) ,
 \frac{1}{q^2} (2{\bf k + q}).{\bf q} \right)
 \langle {\bf k} | \delta \rho (t=0)  |{\bf k} + {\bf q} \rangle        }
\end{array}
\end{equation}
The variation in the energy density can be written in terms of the quantities
(\ref{12}), yielding
\begin{equation} \label{6}
W_n(t)- W_p(t)  = 2 V_0 \delta \rho({\bf r},t)
+ 2 V_1 \nabla .\delta \rho({\bf r},t)  \nabla
+ 2 V_1 \delta \tau({\bf r},t) + 2i
V_1 \left( {\bf \nabla}. \delta {\bf j} + \delta {\bf j}. {\bf \nabla} \right)
\end{equation}
In this formula $V_0$ and $ V_1$ are related to the parameters
of the Skyrme interaction via the following expressions (see appendix A)
\begin{equation} \label{7}\begin{array}{ll}
&{\displaystyle
V_0= - \frac{t_0}{2} \left(x_0+ \frac{1}{2} \right) - \frac{t_3}{12}\left(
x_3 + \frac{1}{2} \right) \rho_0^{\alpha} - \frac{q^2}{16} ( 3 t_1
( 1+2 x_1) + t_2 ( 1+ 2 x_2) )},\\
&{\displaystyle V_1= \frac{1}{16} ( t_2( 1+ 2 x_2) -  t_1( 1 + 2 x_1) )}
\end{array}
\end{equation}
where $\rho_0$ is the equilibrium density of nuclear matter.

The retarded response function is
determined by the corresponding polarizability, i.e., the ratio of the density
change to the field strength
\begin{equation} \label{13}
\Pi(\omega, {\bf q})=\alpha / \varepsilon.
\end{equation}
In the case of the Skyrme effective force, this function is  found  by
noting that
Eq. (\ref{5}) is solved by an Ansatz of the form
\begin{equation} \label{13a}
\delta \rho (t)= \delta \rho (t=0) e^{-i( \omega + i \eta)t},
\end{equation}
provided the matrix elements at time zero satisfy
\begin{equation} \label{14}    {\displaystyle
\langle{\bf k} | \delta \rho (t=0)| {\bf k+q}\rangle =
4 \frac{ f({\bf k+q}) - f({\bf k}) }{ \epsilon({\bf k+q})- \epsilon({\bf k})
+\omega + i\eta }  ( V_0 \alpha +  V_1 {\bf k}.({\bf k+q})\alpha +
\varepsilon +  V_1 \beta -  V_1 \gamma ( 2{\bf k+q}). {\bf q} ) }
\end{equation}
By  multiplying the previous
equation respectively by $1$, $ {\bf k}. ({\bf k+q}) $  and
$ ( 2{\bf k}+{\bf q}). {\bf q} $
and by integrating over ${\bf k}$ we  obtain
the following set of linear equations for $\alpha$, $\beta$ and $\gamma$
\begin{equation} \label{14a}\begin{array}{ll}
&{\displaystyle  \alpha = \left( V_0 + V_1 \Pi_2 \right) \alpha +
V_1 \beta \Pi_0 + 2 m^* \omega
V_1 \Pi_0 \gamma + \epsilon \Pi_0  }\\
&{\displaystyle \beta = \left( V_0 \Pi_2 + V_1 \Pi_4 \right) \alpha +
V_1 \Pi_2 \beta + 2
\gamma m^* \omega V_1 \Pi_2  } \\
&{\displaystyle \gamma = - \frac{2 m^* \omega}{ q^2 \left( 1 - 2
V_1 m^* \rho_0\right) } \alpha }
\end{array}
\end{equation}
The last equation for $ \gamma$ can be checked to be a mere consequense of
the equation of continuity for neutrons:
\begin{equation}
\partial_t \rho_n + \frac{1}{m} {\bf \nabla}. \delta {\bf j} = \frac{1}{2}
( t_1+t_2 ) \left( \rho_n {\bf j} - \rho {\bf j}_n \right)
\end{equation}
with an analogous equation for protons.

Solving the above linear system for $\alpha$ leads to the following expression
of the retarded response function:
\begin{equation} \label{15}
\Pi(\omega, {\bf q}) = \frac{\Pi_{0}
(\omega,{\bf q}) }{1- \bar{V_0} \Pi_{0}(\omega, {\bf q})
- 2 V_1 \Pi_2 (\omega, {\bf q})
- V_1^2 \Pi_2^2 (\omega, {\bf q}) - V_1^2 \Pi_4 \Pi_0  }.
\end{equation}
In this equation  $\Pi_0$ is the unperturbed response function (often
referred to as the Lindhard function \cite{WALECKA}).
In what follows the quantities $\Pi_2$ and  $ \Pi_4$
will be referred to  as generalized Lindhard functions. They are defined as
\begin{equation} \label{16}
\Pi_{2N}(\omega, {\bf q})  = \frac{4}{ (2 \pi)^3}
\int \hbox{d}^3 k  \frac
{ f({\bf k} + {\bf q})- f( {\bf k})  }
{\omega +i \eta-  \epsilon ({\bf k}) + \epsilon({\bf k}+ {\bf q})  }
 \left( {\bf k} .({\bf k +q }) \right) ^N ~,
\end{equation}
where the limit $\eta \to 0^+$ is implicit.
Analytical expressions for the real and imaginary parts of $\Pi_{2N}(\omega,q)$
are given in appendix B as well as the relation between our definitions and
those of other authors \cite{GARCIA}. In Eq. (\ref{15}) we have used a
modified  coefficient $\bar{V_0}$ defined by:
\begin{equation}
 \bar{V_0} =  V_0 - \left( \frac{ m^* \omega}{q} \right)^2 \frac{ 2V_1}
{ 1- 2 V_1 m^*\rho_0 }
\end{equation}
This modified coefficient arises because of the change in the momentum density
induced by the external field.  For interactions with no momentum dependence,
i.e., $t_1=t_2=0$,
we recover the results of Ref.~\cite{BRAVAU}\footnote{ A factor 2 is however
missing in Eq. (13) of this reference, which implies that the interaction
strengths in figs. 1 and 2 of this reference must be divided by a factor 2.}.

Our formula for the response function (\ref{15}) generalizes that of
Garcia- Recio {\it et al}. \cite{GARCIA} to which it reduces at zero
temperature.

\renewcommand{\thesubsection}{B.}
\subsection{ Some properties of the Response Function }

A collective behavior will be observed when the response function exhibits a
peak. Although
this may be the case when the denominator in the response function goes
through a minimum, the most familiar situation corresponds to the case
where there is a pole  in Eq. (\ref{15}). This occurs when the value $\omega_0$
of the frequency is such that
\begin{equation}  \label{16a}
1 = \bar V_0 \Pi_0 + 2 V_1 \Pi_2 + V_1^2 ( \Pi^2_2 + \Pi_0 \Pi_4 )
\end{equation}
The real part of the frequency $\omega_0$ determines the energy of the
resonance and its imaginary part the width. The corresponding
relation is easily
constructed when the imaginary part of $\omega$ is small so that a
linearization of the previous equation near $\Im m (\omega_0)$=0 can be made
\cite{WALECKA}.  In
other cases a numerical construction of the strength function is necessary, as
performed in some of the examples below.

It is interesting to note that the symmetry energy coefficient $a_{\tau}$ of
nuclear matter is, as expected, an important ingredient to determine whether or
not there is a pole. Indeed one has the following relation between $a_{\tau}$
and the coefficients occuring in Eq. (\ref{16a})
\begin{equation} \label{7a}
a_{\tau} = \frac{k_F^2}{6 m^*} + \frac{\rho_0}{2} \left( V_0 +
\frac{q^2}{16} \left( t_2 ( 1+ x_2) + 3 t_1 ( 1+ x_1) \right) +
2 V_1  k_F^2 \right)
\end{equation}
Note that this relation also involves the Fermi momentum
$k_F~=~(3 \pi^2 \rho_0/2)^{1/3}$ and the momentum transfer $q$.

The formula we have derived for the response function looks somewhat cumbersome
and thus it seems worthwhile to explore whether approximations to the previous
scheme can be derived. We have found that one such usefull scheme is provided
by a Thomas- Fermi type appproximation in which one assumes that the change in
the kinetic energy density $\tau$ can be calculated from the proportionality
relation $\tau \sim \rho^{5/3}$, which leads to
\begin{equation} \label{16b}
\delta \tau ({\bf r}, t)/ \tau ({\bf r}, t)= 5 \delta \rho ({\bf r}, t) / 3
\rho ({\bf r}, t).
\end{equation}
Inserting this value into Eq. (\ref{14}) one obtains the approximate
expression for the response function
\begin{equation} \label{16c}
\Pi(\omega, {\bf q}) = \frac{\Pi_{0}
(\omega,{\bf q}) }{1- \tilde{V_0} \Pi_{0}(\omega, {\bf q})
- 2 V_1 \Pi_2 (\omega, {\bf q})},
\end{equation}
in which the coefficient $\tilde V_0$ is defined by
\begin{equation} \label{16d}
\tilde{V_0}= V_0 - 2 V_1 k_F^2.
\end{equation}
This approximate formula was explored in Ref. \cite{RIKEN}. It was found to
reproduce correctly the main features of the exact formula derived in the
previous section: existence of a zero sound at zero temperature with a
disappearance of this mode at temperatures of a few MeV.

In the limit $ \omega \rightarrow 0 $ and $ q \rightarrow 0 $,
the response function becomes the static polarizability.
In the isovector channel, the latter is related to the symmetry energy
coefficient $a_{\tau}$ at temperature $T$. Indeed for $q= \omega=0$ the energy
density of nuclear matter in the presence of a small external field is given by
\begin{equation} \label{16k}
{\cal H}={\cal H}_0+ \frac{a_{\tau}(T)}{\rho_0} (\delta \rho)^2 +
\epsilon \delta \rho = {\cal H}_0 + \frac{a_{\tau}}{\rho_0} \alpha^2 +
\varepsilon \alpha.
\end{equation}
By minimizing with respect to $\alpha$ we find
\begin{equation} \label{16e}
\alpha= - \frac{ \varepsilon \rho_0}{2 a_{\tau}}~,
\end{equation}
yielding
\begin{equation} \label{16f}
\Pi( \omega=0, q=0)= - \frac{ \rho_0}{2 a_{\tau}}.
\end{equation}
This formula can also be obtained by performing the limit
$ \omega \rightarrow 0 $ and $ q \rightarrow 0 $ in the expression of the RPA
response function (\ref{15}), which provides a check of our formula for this
quantity.

The value of the strength distribution per unit volume $S(\omega)$ for the
operator $\tau_3 \exp(i {\bf q}.{\bf r})$  is proportionnal to the imaginary
part of the response function:
\begin{equation} \label{30}
S(\omega)= -\frac{1}{\pi} \Im m\Pi(\omega,q).
\end{equation}
It is also related to the photoabsorbtion strength distribution $S_{abs}$
\cite{CHOMAZ}
\begin{equation} \label{30a}
S_{abs}(\omega)= -\frac{1}{\pi} \frac{1}{1-e^{-\beta \omega}}
\Im m\Pi(\omega,q).
\end{equation}
Furthermore, it satisfies the energy weighted sum rule \cite{CHOMAZ}
\begin{equation} \label{31}
\int^{\infty}_{0} {\rm d} \omega ~\omega S(\omega)
= \frac{q^2 }{2m^*} \rho_0
\left( 1 + \kappa \right)
\end{equation}
where
\begin{equation} \label{31a}
 \kappa = - \frac{ m^* \rho_0}{ 8 } ( t_2 ( 1+ 2 x_2) - t_1
( 1+ 2 x_1)  ~)
\end{equation}
is the enhancement factor arising from the
the momentum dependent terms of the Skyrme interaction.

\setcounter{equation}{0}
\section{ Results and discussion}

The strength function is plotted for the Skyrme forces SGII \cite{SAGI}, SkM
\cite{SKM}, SIII,
SI and SV \cite{VAUBRI,BEFLO} in Figs. 1, 2, 3,  4 and 5
 respectively, as a
function of the excitation energy  $\omega$ for various values of the
temperature. The corresponding values of the Skyrme
parameters are given in appendix A. In these figures we have chosen the value
of the momentum transfer $q$ in such a way that it corresponds to the dipole
mode in lead-208 described by the Steinwedel- Jenssen model \cite{SCHUCK}.
In this model neutrons oscillate against protons inside a sphere of radius $R$
according to the formula
\begin{equation} \label{32}
\langle{\bf r}| \delta \rho|{\bf r}\rangle= \varepsilon
\sin ({\bf q}.{\bf r}) \sin( \omega t),
\end{equation}
where
\begin{equation} \label{33}
q = \frac{\pi}{2 R}.
\end{equation}
Taking $R=6.7$ fm for lead-208 we find $q=0.23$ fm$^{-1}$.

We have checked in our numerical calculations that the energy weighted sum rule
is well satisfied. Some examples are shown in Table 1, which compares the
right hand side of the sum rule [c.f., Eq. (\ref{31})] to the value of the
integrated strength using Simpson's rule.
It can be checked that there is an excellent agreement. There are some
apparent exceptions however which correspond to the presence of a sharp zero
sound peak which falls in between two meshpoints of the integration method. We
have checked that the contribution of the pole just provides the missing
strength in these cases.

For interactions SGII, SkM and SIII
the strength function shows a strong temperature dependence
and exhibits a sharp peak at zero temperature. This peak occurs at a value of
18 MeV which is slightly higher than the observed one (we assume that we
can rely on
the Steinwedel- Jenssen model). One common feature of the three interactions
SGII, SkM and SIII is that their effective masses are very close
(respectively 0.78,  0.79 and 0.76, in units of the bare mass).

\begin{table}[t]
{\footnotesize \caption{
Right hand side (RHS) of the energy weighted sum rule (MeV $\times$ fm$^{-3}$)
compared to the integrated value of the strength $m_1$
for $T$ = 0, 3 and 6  MeV. } }
\centerline {\begin{tabular}{c c c c c c }  \\ \hline \hline
   & RHS  & $m_1(T=0)$ & $m_1(T=3)$ & $m_1(T=6)$ \\ \hline
SGII &  52 & 25  & 52  & 52  \\
SkM & 54   & 15  & 54 & 54 \\
SIII & 49 & 18 & 49  & 49  \\
SI & 43  &  11   & 43  & 43  \\
SV & 121  & 122  & 121 & 120  \\
\hline \hline
\end{tabular} }
\end{table}

For a Skyrme interaction with an effective mass closer to unity, such as SI
(see Fig. 4), the resonance is found at a lower energy (15.5 MeV).
For Skyrme forces with a very small effective mass, such as SV
($m^*/m = 0.36$), the energy of the dipole resonance (see Fig. 5)
is much higher (about 35 MeV) than the observed value
and the dipole strength is nearly independent of temperature.
It still exhibits a weak collective behavior but
for most of the energy range, the form of the response
function is not very different from the
imaginary part of the bare Lindhard function $\Pi_0$, which is shown in
Fig. 6. The imaginary part of
$\Pi_0$ exhibits a nearly linear growth with energy
which is also present in the imaginary part of the RPA response function, with
moderate deviations occuring only in the resonance region.

Let us now try to understand qualitatively the previous results about the
position of the resonance and its evolution with temperature. The imaginary
part of the bare response function $\Pi_0$ is known to show at zero
temperatures two angular points at the following values of the energy
(see appendix B and Fig. 6)
\begin{equation} \label{33a}
\displaystyle
\omega_{\pm} = \frac{ q k_F}{m^*} \pm \frac{q^2}{2 m^*}.
\end{equation}
As temperature rises the corresponding discontinuities are smeared by the
presence of the Fermi occupation numbers (Fig. 6). At these points the real
part of the bare response function has a vertical slope (at zero temperature)
and a maximum in between these points. For the particular value of the
momentum transfer we are considering ($q$=0.23 fm$^{-1}$) and for a Fermi
momentum $k_F$=1.36 fm$^{-1}$ one finds
\begin{equation} \label{33b}
\omega_{\pm}=
\frac{m}{m^*} \times (15 \pm 1.5) {\rm MeV}.
\end{equation}
Let us now assume for more simplicity
(which is supported by numerical estimates) that the Lindhard function of order
zero is the most important one for our discussion, so that the RPA response
(\ref{15}) can be approximated as
\begin{equation} \label{33c}
\Pi(\omega, {\bf q}) \simeq \frac{\Pi_{0}
(\omega,{\bf q}) }{1- V_0 \Pi_{0}(\omega, {\bf q}) }.
\end{equation}
At this point we should  note that the maximum value of the real part of the
bare response is proportional to the effective mass
\begin{equation} \label{33f}
\max (\Re e \Pi_0) = m^* k_F A(k_F, q, T),
\end{equation}
as can be seen from the equations given in appendix B.
Since the maximum is smeared and reduced at high temperatures
the function $A$ decreases as temperature increases.
In terms of this function the condition for
the existence of a zero sound is
\begin{equation} \label{34}
1 \leq V_0 m^* k_F A(T).
\end{equation}
It turns out
that for interactions with an effective mass close to unity and at zero
temperature the condition (\ref{34}) is just satisfied. If this is the case,
we expect the response function to have a maximum when the
real part of the bare response is near its maximum, i.e., between the two
points defined by Eq. (\ref{33a}). Since these points are quite close, a
reasonable estimate of the resonance energy is
\begin{equation} \label{33d}
\omega_{R}\simeq \frac{m}{m^*} \times 15  {\rm MeV}.
\end{equation}
By comparing this formula with the results in Figs. 1-5, it can be seen to
provide a good description of the resonance energies. Note that for other
values of the momentum transfer $q$, a similar construction would give the
following dispersion relation
\begin{equation} \label{33e}
\omega_{R} \simeq  c_0 \times q,
\end{equation}
with
\begin{equation} \label{35}
c_0 \simeq \frac{k_F}{m^*}=v_F= 0.3 \times c \times \frac{m}{m^*},
\end{equation}
where we have taken $k_F$=1.36 fm$^{-1}$. This formula
exhibits the important role played by the value of the effective mass.
In actual calculations the sound velocity $c_0$ is slightly larger than the
Fermi velocity as can be seen from Table 2. The similarity between the two
velocities shows that nuclear matter exhibits weaker collective effects than
helium- 3. Indeed, in this last case the zero sound
velocity is more than 3 times the Fermi velocity \cite{NEGELE}.

The dispersion relation $\omega_R= c_0 q$ implies that the resonance in light
nuclei is located higher in energy according to the relation $q=2 \pi / R$,
i.e., evolves with mass number as $\omega_R \sim A^{-1/3}$ in agreement with
the empirical formula \cite{SCHUCK}. Since in our model the strength is
concentrated in a single region this result also means
weaker collectivity in light nuclei.

\begin{table}[t]
{\footnotesize \caption{
Velocity $c_0$ of the collective mode
compared to the Fermi velocity  $v_F$  for the Skyrme forces SI, SGII
and SV.}}
\centerline {\begin{tabular}{c c c c  c } \hline \hline
   & SI  & SV   & SGII \\ \hline
$c_0$ & 0.35  & 0.77  & 0.40 \\
$v_F$ & 0.30  & 0.77  & 0.35 \\  \hline  \hline
\end{tabular} }
\end{table}

Let us now show that the previous formulae also provide an explanation for
the presence or not of a zero sound at zero temperature and for
the different temperature dependences of the strength obtained for various
interactions.
Indeed from the relation between $V_0$, $a_{\tau}$ and $m^*$ we have
\begin{equation} \label{36}
\frac{m^*}{m} \frac{V_0}{2} \rho_0= \frac{m^*}{m} a_{\tau} - \frac{k_F^2}{6m},
\end{equation}
Taking $a_{\tau}$=30 MeV and $k_F$= 1.36 fm$^{-1}$ this relation shows that the
quantity $m^* V_0$, which is the relevant one for our discussion,
 is much smaller for interactions with a small effective mass
(0.4) than for interactions with $m^*/m \simeq 1$. Therefore
if the condition (\ref{34}) is just satisfied at zero temperature for
$m^*/m$=1, such will not be the case for an interaction with $m^*/m$=0.4.
Similarly, since the function $A$ decreases with temperature it is also clear
that the condition for the existence of a zero sound will eventually no
longer hold at high enough temperatures (in actual calculations a few MeV).

\section{Conclusion}

In conclusion we have found that the standard Skyrme forces SGII, SII and SkM
give rise at zero temperature to a zero sound type collective mode exhibiting
the usual dispersion relation  $ E_{\rm res} = c_0 q $.
The sound velocity for these forces is just slightly greater than the Fermi
velocity, which implies a rather weak collective behaviour as compared to
helium-3 for which there is a factor three between the two velocities.
For values of the
momentum transfer corresponding in the Steinwedel-Jenssen model to the giant
dipole mode in lead-208, we have found that the zero sound mode disappears at
temperatures of a few MeV. This may be related to the saturation of photon
multiplicities observed in some recent experiments \cite{LEFAOU}. For Skyrme
forces with a small value of the effective mass such as SV, we have found no
zero sound at zero temperature and a weak variation of the strength with
temperature. It is worthwhile noting that the previous forces all provide good
descriptions of nuclear properties, such as binding energies and radii, all
over the periodic table. In spite of this common property, they do give rather
different predictions for the temperature evolution of giant resonances and
also for their positions. Collective properties thus appear to give useful
information on the effective nucleon- nucleon interaction.

One limitation of our calculations is that although they do include the effect
of the volume symmetry energy, they ignore the effect of the surface symmetry
energy wich is known to play a role, especially in light nuclei \cite{JENJAC}.
They ignore as well shell effects which are also important in light nuclei.
These effects indeed produce in this case a fragmentation of the strength in
RPA
calculations (see for instance the results of Sagawa and Bertsch \cite{SAGAWA}
using the SGII force) whereas our model produces only a broad peak.
Complete RPA calculations in finite nuclei (including heavy nuclei)
would thus be of interest to complete
the results of the present work. For such calculations we believe that the
discussion we have presented would be a usefull guide.

\vskip 0.3cm
\noindent {\Large {\bf Acknowledgements}}
\vskip 0.2cm
We are grateful to Nguyen Van Giai and Nicole Vinh Mau for stimulating
discussions. The Division de Physique Th\'eorique is ``Unit\'e de Recherche des
Universit\'es Paris XI et Paris  VI associ\'ee au CNRS''.

\vskip 0.5 cm
\setcounter{equation}{0}
\renewcommand{\theequation}{A.\arabic{equation}}
\noindent {\Large {\bf Appendix A: Parameters of the interactions used} }
\vskip 0.2cm

We consider an effective Skyrme interaction of the following form:
\begin{equation} \label{2a} \begin{array}{ll}
v_{12} = &{\displaystyle
 t_0 ( 1 + x_0 P_{\sigma} ) \delta ({\bf r_1} - {\bf r_2}) +
\frac{t_1}{2} \left(1+ x_1 P_{\sigma} \right) \left[ \delta({\bf r_1} -
{\bf r_2}) k^2 + k'^2 \delta (
{\bf r_1} - {\bf r_2}) \right]  + }\\
&{\displaystyle
t_2 \left( 1+ x_2 P_{\sigma} \right) {\bf k'}. \delta ({\bf r_1} - {\bf r_2})
{\bf k}  + \frac{t_3}{6} ( 1 + x_3 P_{\sigma} ) \rho^{\alpha} \delta
({\bf r_1} - {\bf r_2})}
\end{array}
\end{equation}
where $P_{\sigma}$ is the spin
exchange operator.

Denoting the single particle wave functions by
$\phi_i ({\bf r} ,\sigma, q) $, $\sigma $ and $q$
being the labels for spin and isospin, the nucleon density
$\rho_q ({\bf r})$, the kinetic energy density $ \tau_q ({\bf r})$ and
the momentum density
$ {\bf j}({\bf r})$  can be expressed as
\begin{equation} \label{3} \begin{array}{ll}
\rho_q ({\bf r}) & = \sum_{i,\sigma} |\phi_i({\bf r}, \sigma, q)|^2   \\
\tau_q ({\bf r}) & = \sum_{i,\sigma} |\nabla \phi_i({\bf r}, \sigma, q)|^2 \\
{\bf j} ({\bf r}) & = \sum_{i,\sigma} \frac{1}{2i}
\left( \nabla - \nabla ' \right)
\phi^*_i ({\bf r} ',\sigma,q) \phi_i ({\bf r} ,\sigma,q)
\vert_{{\bf r} = {\bf r} '}
\end{array}
\end{equation}
\begin{table}[t]
{\footnotesize \caption{
 Numerical values of the  parameters
$t_0$, $t_1$, $t_2$, $t_3$, $x_0$, $x_1$, $x_2$, $x_3$ and  $\alpha$
corresponding to the five Skyrme interactions SGII, SkM, SIII, SI and SV
considered in this work.}
}
\begin{tabular}{c c c c c c c c c c}   \\ \hline \hline
&$t_0$&$t_1$&$t_2$&$t_3$&$x_0$&$x_1$&$x_2$&$x_3$&$\alpha$ \\
   & (MeV fm$^3$)&( MeV fm$^5$)&(MeV fm$^5$)&(MeV fm$^6$)& & & & &\\ \hline
SGII$^a$& -2645.&340.&-41.9&15595.&.09&-.588&1.425&.0604&1/6 \\
SkM$^b$& -2645.&410.&-135.&15595.&.09& 0&0 &0 & 1/6 \\
SIII$^c$ & -1128.75 & 395. & -95.0 & 14000. & .45 &0 &0& 1&1 \\
SI$^c$ & -1057.3 & 235.9 & -100.0 & 14463.5 & .56&0 & 0&1&1 \\
SV$^c$&  -1248.29 & 970.56 & 107.22  & 0. & -.17 &0 &0&1& 1 \\
\hline \hline \\
\end{tabular}

$^a$  From Ref. \cite{SAGI}. $^b$ From Ref. \cite{SKM}, $^c$ From Refs.
\cite{VAUBRI,BEFLO}.
\end{table}
The Hartree-Fock mean field Hamiltonian reads:
\begin{equation} \label{3a}
W_q ({\bf r} )= {\bf \nabla} \frac{1}{2m^*_q} {\bf \nabla}  + U_q(r),
\end{equation}
where $U_q$ is a local potential given in the case of neutrons (q = n) by:
\begin{equation} \label{4} \begin{array}{ll}
U_n({\bf r},t) = & t_0 \{ \left( 1 + x_0\frac{1}{2} \right) \rho({\bf r},t) -
\left( x_0 + \frac{1}{2} \right) \rho_n({\bf r},t) \}  \\
& + t_3\frac{1}{12} \left\{ (2+\alpha) \left( 1 + \frac{x_3}{2} \right)
\rho^{(\alpha+1)} ({\bf r},t) ) \right\} \\
  & - t_3 \{
\left( x_3 +\frac{1}{2} \right) ( 2 \rho_n({\bf r},t) \rho^{\alpha}
({\bf r},t) +
\alpha ( \rho_p^2({\bf r},t) + \rho_n^2({\bf r},t) ) \rho^{(\alpha -1)}  \} \\
& + \{ \frac{1}{4}t_1 ( 1+ \frac{x_1}{2} )  +
\frac{1}{4} t_2 (1+ \frac{x_2}{2} ) \} \left( \tau - \frac{1}{2i}
\left({\bf \nabla}.{\bf j} + {\bf
j}.{\bf \nabla} \right) \right) \\
& + \{ - \frac{1}{4} t_1 ( x_1 +\frac{1}{2} ) + \frac{1}{4} t_2( x_2 +
\frac{1}{2} )  \} \left( \tau_n - \frac{1}{2i}
\left({\bf \nabla}.{\bf j}_q + {\bf j}_q.{\bf \nabla} \right) \right) \\
&  + \left\{ - \frac{3}{8}t_1 (\frac{1}{2} +x_1)  + \frac{1}{8} ( \frac{1}{2} +
x_2) \right\} \nabla^2 \rho
 + \left\{ \frac{3}{8} t_1 ( \frac{1}{2} + x_1) +
\frac{1}{4} t_2 ( \frac{1}{2} + x_2 ) \right\} \nabla^2 \rho_n
\end{array}
\end{equation}
with a similar expression for protons.
The effective mass $ m^*_q$ is given by:
\begin{equation}
\frac{1}{2 m^*_q} = \frac{1}{2 m} + \frac{1}{4} \left(t_1 (1+
\frac{x_1}{2} ) + t_2 ( 1 + \frac{x_2}{2})
\right) \rho + \frac{ 1}{8} \left( t_2 ( x_2 + \frac{1}{2})  - t_1 ( x_1 +
\frac{1}{2} )  \right)\rho_q
\end{equation}

\vskip 0.5 cm
\setcounter{equation}{0}
\renewcommand{\theequation}{B.\arabic{equation}}
\noindent {\Large {\bf Appendix B: Expression of the response function}}
\vskip 0.2cm

In this appendix we give the explicit expressions of the real and imaginary
parts of the generalized Lindhard functions defined in Eqs. (\ref{16}).
For this purpose we need to define the following integrals:
\begin{equation} \label{18}
I_{2N} = \frac{2}{(2 \pi)^3} \int {\rm d}^3 k k^{(2N)}
\frac{ \left( f({\bf k+q}) -
f({\bf k}) \right)}{ \omega + i \eta + \epsilon ({\bf k+q}) - \epsilon({\bf k})
}
\end{equation}
The imaginary parts of the generalized Lindhard functions are given by
\begin{equation} \label{19} \begin{array}{ll}
 \Im m \Pi_{0}(\omega, {\bf q})  = &{\displaystyle -\frac{m^{*2}}{\pi q\beta}
 \log \frac{1+ e^{\beta (\mu-E_- )}}{1+e^{\beta(\mu-E_+)}} } \\
 \Im m \Pi_2 (\omega, {\bf q})  = &{\displaystyle
-\frac{2m^{*3} }{\pi \beta^2 q} \left(\beta
\sqrt{E_{+} E_{-}} \log \frac{1+ e^{\beta (\mu-E_- )}}
{1+e^{\beta(\mu-E_{+})} } +
{\rm Li}_2 \left(1+e^{\beta(\mu - E_{+})}\right) - {\rm Li}_2
\left( 1+ e^{\beta(\mu - E_{-}) } \right)   \right) } \\
 \Im m \Pi_4 (\omega, {\bf q})  = &{\displaystyle
- q \sqrt{2m^* E_+} \Im m \Pi_2
+ 2 \Im m I_4 - 2 \sqrt{2 m^* E_+}
q \Im m I_2 }
\end{array}
\end{equation}
where ${\rm Li}_2$ is the Euler dilogarithmic function \cite{ABRAMOWITZ}
\begin{equation} \label{29}
{\rm Li}_2(x)= \int^{x}_1 \frac{ \log (t)}{t- 1} {\rm d} t
\end{equation}
and
\begin{equation} \label{20}
E_{\pm}=\frac{m^*}{2 q^2} (\omega \pm \frac{q^2}{2 m^*})^2.
\end{equation}
The expressions of the functions  $ \Im m I_2$ and $ \Im m I_4$ are :
\begin{equation} \label{28}\begin{array}{ll}
{\displaystyle \Im m I_2(\omega , q) = -\frac{m^{*3}}{\pi q \beta^2}
\left( \beta E_+ \log \frac{1+ e^{\beta (\mu -E_-)}}
{1+e^{\beta(\mu - E_+)}} +{\rm Li}_2(1+ e^{\beta (\mu -E_+)}) -
{\rm Li}_2(1+ e^{\beta (\mu -E_-)}) \right) } \\
{\displaystyle \Im m I_4(\omega , q) = -\frac{2 m^{*4}}{ \pi q \beta }
E_+^2 \left(
\log \frac{1+ e^{\beta (\mu -E_-)}}{1+e^{\beta(\mu - E_+)}} +
2 \int^{\infty}_{1} {\rm d}z ~z  \log \frac{1 + e^{-\beta(z E_+ - \omega
- \mu )} }{ 1 + e^{ -\beta(z E_+ - \mu) } } \right) }
\end{array}
\end{equation}
At zero temperature the real parts of the $\Pi_{2N}$ are given by
\begin{equation} \label{21} \begin{array}{ll}
 \Re e \Pi_{0} (T=0)  = &{\displaystyle
\frac{m^*k_F}{\pi^{2}} \left( -1 + \frac{k_F}{2q}\left[ \phi(x_+)+
\phi(x_-)\right] \right)}  \\
 \Re e \Pi_2 (T=0) = &{\displaystyle \frac{m^* k_F^3 }{2 \pi^2} \left( -3
+ x_+ x_- + x_+^2 + x_-^2 +\right.}\\&{\displaystyle ~
 \left. \frac{k_F}{2q} \left[ (1- x_+^2-2x_+x_-) \phi(x_+)
+(1-x_-^2-2x_+x_- ) \phi(x_-)
\right] \right) } \\
 \Re e \Pi_4 (T=0) = &{\displaystyle 2 \left( \Re e I_4(T=0) -
2q \sqrt{2 m^*E_+ } \Re e I_2(T=0)~+ \right.}\\&{\displaystyle ~~~~~~~~~~
\left. q^2  m^*E_+ \Re e \Pi_0 (T=0)+
\frac{1}{3\pi^2}m^*q^2k_F^3  \right) } \end{array} \end{equation}
where
\begin{equation} \label{22} \begin{array}{ll}
 \Re  e I_2 (T=0) = &{\displaystyle \frac{m^* k_F^3 }{4 \pi^2} \left(
-3 -x_+x_-- x_+^2 + x_-^2 + \frac{k_F}{2q} \left[ (1+ x_+^2) \phi(x_+)
+(1+x_-^2+ \frac{4m^* \omega}{k_F^2} ) \phi(x_-)
\right] \right) } \\
 \Re e I_4 (T=0) = &{\displaystyle -\frac{m^* k_F^6}{2 \pi^2 q} \left( ~(
\frac{1}{6} ( 1 +x_-^2 + x_-^4 ) + \frac{m^* \omega}{k_F^2} (1 +x_-^2) +
\frac{2m^{*2} \omega^2}{k_F^4} ) \phi(x_-) + \right.}\\
&{\displaystyle
\frac{1}{6} ( 1 +x_+^2 + x_+^4 ) \phi(x_+) +  \frac{5 q}{3 k_F} + \frac{1}{3}
( \frac{1}{3} x_-^3 + \frac{1}{3} x_+^3 + \frac{2 m^* \omega x_-}{k_F^2} +
\frac{ 8 q^2 x_+}{k_F^2} ) +}\\
&{\displaystyle  \left. \frac{x_-^5}{3} + \frac{x_+^5}{3} + \frac{2
m^* \omega x_-^3}{k_F^2} + \frac{4 m^{*2} \omega^2 x_-}{k_F^4} \right)}
\end{array}
\end{equation}
with
$\displaystyle
x_{\pm}= \frac{q}{2k_F} \pm  \frac{m^* \omega}{ qk_F },
$~~
and $\displaystyle \phi (x)= (1-x^2) \log | \frac{x-1}{x+1} |$.

For non zero temperature the expression of
$\Re e \Pi_{2N}$ ($N$=0,1,2) is an average of the zero temperature functions
calculated for the same values of $\omega$ and $q$, but with various values of
the Fermi momentum $k_F$ distributed with a weight factor which is just the
derivative of the Fermi occupation number. Explicitly one has the following
formula:
\begin{equation} \label{25} \begin{array}{ll}
\Re e\Pi_{0}(\omega,q,T)= -\int
\Re e\Pi_{0}(\omega,q,T=0,k_F=k) ~\hbox{d} f(k,T) \\
\Re e\Pi_{2}(\omega,q,T)= -\int
\Re e\Pi_{2}(\omega,q,T=0,k_F=k) ~\hbox{d} f(k,T)\\
\Re e\Pi_{4}(\omega,q,T)= -\int
\Re e\Pi_{4}(\omega,q,T=0,k_F=k) ~\hbox{d} f(k,T),
\end{array}
\end{equation}
where $f(k,T)$ is the occupation number. For the case of zero temperature we
have $$ {\rm d} f(k)~=~-~\delta (k-k_F)~{\rm d}k, $$ yielding the above
expressions for these functions.

We would like also to show the relation between our definition of generalized
Lindhard functions ( $ \Pi_{2i}$ ) and those of Garcia- Recio {\it et al.}
\cite{GARCIA} ($ \bar \Pi_{2N}$) defined in the limit $ T=0$
\begin{equation} \begin{array}{ll}
{\displaystyle \Pi_0 = 4 \bar \Pi_0 } \\
{\displaystyle \Pi_2 = 4 \bar \Pi_2 - 2 q^2 \bar \Pi_0 } \\
{\displaystyle \Pi_4 = 4 \bar \Pi_4 - 4 q^2 \bar \Pi_2 - q^2 m^* \rho_0 + q^4
\bar \Pi_0 + (2 m^* \omega q )^2 \bar \Pi_0 }
\end{array}
\end{equation}
Our definition of $ V_0 $ and  $ V_1$ and their $ W_1$  and  $ W_2$  are
related by
\begin{equation} \begin{array}{ll}
{\displaystyle V_0  = \frac{W_1}{4} + \frac{q^2}{8} W_2 } \\
{\displaystyle V_1  = \frac{W_2}{4} }
\end{array}
\end{equation}

\vspace{0.7cm}
\noindent {\Large {\bf  Figure captions}}

\vskip 0.5cm
\noindent{\bf Figure 1} Distribution of strength per unit volume for the
operator
$\exp (i {\bf q}. {\bf r})$ (in fm$^{-2}$) as a function
of the energy $\omega$ (in MeV) for a momentum $q=0.23$ fm$^{-1}$ and for
different values of the temperature $T=0\to 6$ MeV, in the case of the
interaction SGII.

\noindent{\bf Figure 2} Same as Fig. 1 for the force SkM.

\noindent{\bf Figure 3} Same as Fig. 1 for the force SIII.

\noindent{\bf Figure 4} Same as Fig. 1 for the force SI.

\noindent{\bf Figure 5} Same as Fig. 1 for the force SV.

\noindent{\bf Figure 6} Imaginary part of $\Pi_0$ ( in fm$^{-2}$)  as a
function of the energy $\omega$ ( in MeV), in the case of SV force,
for $q = 0.23$ fm$^{-1}$; the temperatures are from $T$=0 to $T$=6 MeV.

\end{document}